\documentclass[conference]{IEEEtran}
\IEEEoverridecommandlockouts
\usepackage{cite}
\usepackage{url}
\usepackage{amsmath,amssymb,amsfonts}
\usepackage{algorithmic}
\usepackage{graphicx}
\usepackage{textcomp}
\usepackage{xcolor}
\def\BibTeX{{\rm B\kern-.05em{\sc i\kern-.025em b}\kern-.08em
    T\kern-.1667em\lower.7ex\hbox{E}\kern-.125emX}}

\usepackage[ruled,vlined]{algorithm2e}
\usepackage{booktabs}

\usepackage{amsthm}

\begin{document}

%\title{An Energy-Efficient Framework for joint Routing, Radio Resource Management and Task Offloading for Earth surveillance with LEO Constellations\\
\title{Edge Computing and Communication for Energy-Efficient Earth Surveillance with LEO Satellites\\
%{\footnotesize \textsuperscript{*}Note: Sub-titles are not captured in Xplore and should not be used}
\thanks{Part of the research has been supported by the project SatNEx-V, co-funded by the European Space Agency (ESA). This work has also received funding by the Spanish ministry of science and innovation under project IRENE (PID2020-115323RB-C31 / AEI / 10.13039/501100011033) and grant from the Spanish ministry of economic affairs and digital transformation and of the European union – NextGenerationEU [UNICO-5G I+D/AROMA3D-Space (TSI-063000-2021-70).}
}

\author{\IEEEauthorblockN{Marc Martinez-Gost\IEEEauthorrefmark{1}\IEEEauthorrefmark{2},
Israel Leyva-Mayorga\IEEEauthorrefmark{3}, Ana Pérez-Neira\IEEEauthorrefmark{1}\IEEEauthorrefmark{2}, \\ Miguel Ángel Vázquez\IEEEauthorrefmark{1}, Beatriz Soret\IEEEauthorrefmark{3}\IEEEauthorrefmark{5},  Marco Moretti\IEEEauthorrefmark{4}}
\IEEEauthorblockA{
\IEEEauthorrefmark{1}Centre Tecnològic de Telecomunicacions de Catalunya, Spain\\
\IEEEauthorrefmark{2}Dept. of Signal Theory and Communications, Universitat Politècnica de Catalunya, Spain\\
\IEEEauthorrefmark{3}Dept. of Electronic Systems, Aalborg University, Denmark\\
\IEEEauthorrefmark{4}Dept. of Information Engineering, University of Pisa, Italy\\
\IEEEauthorrefmark{5}Dept. of Communications Engineering, University of Málaga, Spain\\
Emails: \{mmartinez, aperez, mavazquez\}@cttc.es,\\
\{ilm, bsa\}@es.aau.dk, marco.moretti@unipi.it}
}

% \author{\IEEEauthorblockN{Marc M. Gost}
% \IEEEauthorblockA{\textit{Dept. of Signal Processing and Communications} \\
% \textit{Universitat Politecnica de Catalunya}\\
% Barcelona, Spain \\
% marc.martinez.gost@estudiantat.upc.edu}
% \and
% \IEEEauthorblockN{Ana Perez-Neira}
% \IEEEauthorblockA{\textit{Dept. of Signal Processing and Communications} \\
% \textit{Universitat Politecnica de Catalunya}\\
% Barcelona, Spain \\
% ana.isabel.perez@upc.edu}
% \and
% \IEEEauthorblockN{Miguel Ángel Vázquez}
% \IEEEauthorblockA{\textit{Centre Tecnologic de}\\ \textit{Telecomunicacions de Catalunya} \\
% Castelldefels, Spain \\
% mavazquez@cttc.es}
% \and
% \IEEEauthorblockN{Israel Leyva-Mayorga}
% \IEEEauthorblockA{\textit{Department of Electronic Systems} \\
% \textit{Aalborg University}\\
% Aalborg, Denmark \\
% ilm@es.aau.dk}
% \and
% \IEEEauthorblockN{Beatriz Soret}
% \IEEEauthorblockA{\textit{Department of Electronic Systems} \\
% \textit{Aalborg University}\\
% Aalborg, Denmark \\
% bsa@es.aau.dk}
% \and
% \IEEEauthorblockN{Petar Popovski}
% \IEEEauthorblockA{\textit{Department of Electronic Systems} \\
% \textit{Aalborg University}\\
% Aalborg, Denmark \\
% petarp@es.aau.dk}
% \and
% \IEEEauthorblockN{Marco Moretti}
% \IEEEauthorblockA{\textit{Dept. of Information Engineering} \\
% \textit{University of Pisa}\\
% Pisa, Italy \\
% marco.moretti@unipi.it}
% }

\maketitle

\begin{abstract}

Modern satellites deployed in low Earth orbit (LEO) accommodate processing payloads that can be exploited for edge computing. Furthermore, by implementing inter-satellite links, the LEO satellites in a constellation can route the data end-to-end (E2E). These capabilities can be exploited to greatly improve the current store-and-forward approaches in Earth surveillance systems. However, they give rise to an NP-hard problem of joint communication and edge computing resource management (RM). In this paper, we propose an algorithm that allows the satellites to select between computing the tasks at the edge or at a cloud server and to allocate an adequate power for communication. The overall objective is to minimize the energy consumption at the satellites while fulfilling specific service E2E latency constraints for the computing tasks. Experimental results show that our algorithm achieves energy savings of up to 18\% when compared to the selected benchmarks with either 1) fixed edge computing decisions or 2) maximum power allocation.

\end{abstract}

%—Mobile edge computing (MEC) can efficiently minimize computational latency, reduce response time, and improve quality-of-service (QoS) by offloading tasks in the access network. Although lots of edge computation offloading schemes have been proposed in terrestrial networks, the hybrid satellite terrestrial communication, as an emerging trend for the next generation communication, has not taken edge computing into consideration.

%\begin{IEEEkeywords}
%Edge computing, energy efficiency, LEO satellites, resource management, routing.
%\end{IEEEkeywords}

\section{Introduction}
The last two decades have seen an unprecedented growing trend towards space-based Internet services and the deployment of mega-constellations of LEO satellites by high-tech competitors.
%, such as SpaceX Starlink, OneWeb and Amazon Kuiper \cite{6g_}.
There is a demanding need to address the standardization of the satellite segment with respect to the ground infrastructure, which will play a pivotal role on the path to 6G \cite{6g}.
Extending the legacy 5G NTN use cases of no-served and under-served areas, aviation and maritime use cases, the 6G NTN is meant to gather an extensive number of additional use cases including bulk download of Earth Observation data \cite{white_paper}.
%Opposed to a wired network, a satellite infrastructure provides robustness and more durability.
Delay-sensitive Earth Observation applications are of significant interest, including emergency communications and real-time surveillance. This is the case of the PAZ satellite mission, where satellites take images in which to detect vessels under unauthorized activities \cite{paz}.
%This notion of a satellite constellation for Earth Observation was promoted more than 10 years ago. %\cite{first_leo}
%This is is the case of Disaster Monitoring Constellation (DMC) \cite{dmc}, which was composed of 5 micro-satellites capable of providing multi-spectral images in less than one day with a relevant resolution.
Motivated by the severe reduction of the CAPEX, multi-purpose satellite missions may be launched by combining sensing and communication applications.
%In this context, space segment challenges will show up when accommodating on-board instrument and communication subsystems.
%Currently, Planet acquired RapidEye and it is leading the market of Earth Observation with small satellites missions \cite{planet}. Yet another relevant actor of non-geostationary (NGEO) Earth Observation is the Finish company IceEye \cite{iceye}.

Also, modern satellites accommodate processing payloads, which can improve the surveillance services. In this respect, there are works in Mobile Edge Computing (MEC) that exploit the processing capabilities of the satellite segment beyond a simple relay system. In these works, the data is generated on the Earth and the satellite network accepts computing tasks from ground devices (i.e., tasks are offloaded). For instance, in \cite{sat_iot} an offloading strategy is developed for tasks generated in IoT terrestrial devices. The communication and computation resources are optimized to reduce the latency and power consumption of the satellites. In \cite{hybrid} the terrestrial task can also be offloaded to the cloud server and the objective is to reduce the energy consumption of the ground users subject to coverage time constraints of the LEO. In \cite{double_edge}, the authors extend the previous architectures by allowing each satellite to offload the task up to four more satellites. We note that a key driver is energy consumption. Specifically, LEO satellites have a stringent power constraint as batteries are charged with solar panels and energy-efficient mechanisms will extend their lifespan.

%A key driver in the development of future wireless networks is energy consumption. Specifically, LEO satellites have a stringent power constraint as batteries are charged with solar panels and energy-efficient mechanisms will extend their lifespan. The Mobile Edge Computing (MEC) framework allows to exploit the processing capabilities of the satellite segment beyond a simple relay system. The data is generated on the Earth and the satellite network accepts computing tasks from ground devices (i.e., tasks are offloaded). In \cite{sat_iot} an offloading strategy is developed for tasks generated in IoT terrestrial devices. The communication and computation resources are optimized to reduce the latency and power consumption of the satellites. In \cite{hybrid} the terrestrial task can also be offloaded to the cloud server. The objective is to reduce the energy consumption of the ground users subject to coverage time constraints of the LEO. The authors in \cite{double_edge} minimize the energy consumption under delay constraints. They extend the previous architectures by allowing each satellite to offload the task up to four more satellites.

The previous works assume the terrestrial terminals to have the satellite within Line of Sight (LoS). This simple architecture is assumable when the task is originated on the ground, but not when it comes from the satellite (e.g., processing satellite imagery). In this case, the satellite may not have LoS with any ground station (GS). Current deployments do not route the data through inter-satellite links (ISL), but use a store-and-forward strategy until the satellite has LoS with a GS. This represents a drawback for delay-sensitive services as having visibility of a GS may take up to one day. This promotes the development of RM algorithms for LEO constellations, which are based on inter-satellite routes, to provide shorter latency, better Quality of Service (QoS) and, ultimately, with the satellite segment becoming less dependent of the terrestrial network.
%Along this, the authors in \cite{israel} propose a methodology for establishing dynamic ISLs that maximize the throughput of the LEO network.
%One of the greatest challenges of LEO constellations is the dynamism of the network associated to high speeds.

To the best of our knowledge, there is no work devoted to the decision of where to process a task created at satellite $n$ (satellite edge or GS cloud), while considering radio resource allocation and routing.
Our contribution resides in considering a more realistic and complex architecture of the LEO constellation and to generalize the downlink (DL) problem assuming that $k$ hops are needed to reach the ground from the source LEO. Besides, we adopt a more general power consumption model dedicated to the effect of the power amplifier module. This is usually omitted in the literature and brings critical implications in the optimization.

In this work we consider the problem of jointly optimizing the routes, the transmission powers and the use of the computing resources in order to reduce the energy consumption of the LEO constellation and meet latency constraints.
Since the optimization problem is non-convex and NP, we decouple it into two subproblems: first, the routing procedure via minimization of the propagation time; then, the allocation of transmission power and computing resources are reformulated into a fractional program that provides minimum energy with respect to the preset paths. We call this approach Sat2C.

The main goal is to obtain a first understanding of the different resources' roles: satellites in the route, communication power, CPU processing and computing decisions. The solution to this problem is compared with the store-and-forward baseline algorithm and selected benchmarks. Our approach provides a suitable trade-off between energy consumption and latency.

\section{System Model}
We consider a LEO satellite constellation and a set of GSs acting as cloud servers.
The satellites generate \emph{tasks}, which are blocks of Earth surveillance data that must be processed and stored in the cloud servers (i.e., ground infrastructure) within a pre-defined time window. In such scenario, we investigate the problem of minimizing the overall energy consumption due to computing and communication. The joint optimization of computing and communication resources in this scenario involves:

\emph{1. Routing:} The optimal route towards a nearby GS must be selected.

\emph{2. Edge or Cloud Computing:} The satellites must be able to make an optimal choice between: 1) Edge computing: processing the task locally and then route to the GS and 2) Cloud computing: route the generated data and process the task at the GS.

\emph{3. Radio Resource Management:} Define the optimal power allocation for the ISLs in the selected route.

\begin{figure}[t]
%\vspace*{-15 pt}
\hspace*{-0. cm}
\includegraphics[width=1\columnwidth]{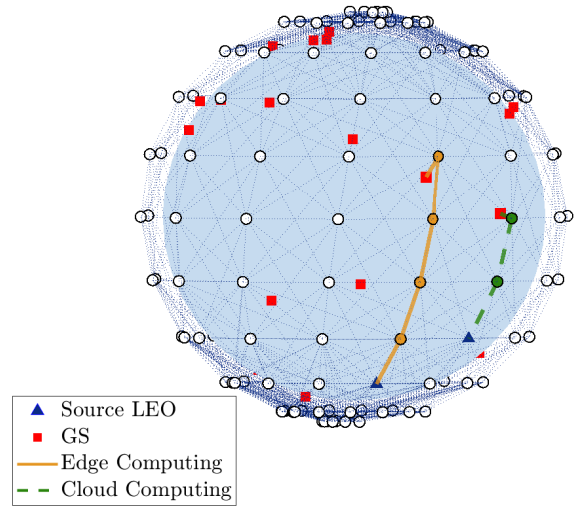}
\centering
\vspace*{-22.5 pt}
\caption{An example of the joint optimization of computing and communication resources for two tasks generated at the LEO satellites. For long routes (yellow), computing the task at the edge minimizes the energy consumption. For short routes (green), computing at the cloud is preferred.}
\vspace*{-0 pt}
\label{fig: scenario}
\end{figure}

Figure \ref{fig: scenario} depicts the result of the optimization for two tasks generated at different satellites. Here, it can be seen that the computing decision depends on the position of the satellites and GSs and, hence, on the length of the route.

While the constellation is dynamic, the satellites have orbital periods of around $90$ minutes. Therefore, the time scale for communication and computation is much shorter than the changes in the network topology. Thus, we assume that the constellation topology is static during the time it takes to complete each task and, hence, observe the constellation at specific time instants. Building on this, we represent the LEO constellation and the GSs at a given time instant as the weighted undirected graph $\mathcal{G} = \left\{\mathcal{V}, \mathcal{E} \right\}$. Specifically, we consider the vertex set $\mathcal{V}$, where $\mathcal{N} \subset\mathcal{V}$ s.t. $|\mathcal{N}|=N$ is the set of LEO satellites generating a task at the specific time instant and $\mathcal{M}=\left\{m\right\}$ s.t. $\mathcal{M}\cap\mathcal{N}=\emptyset$ is the set of GSs. The weighted edge set is $\mathcal{E}$. The deadline to store the result of the task generated by satellite $n\in\mathcal{N}$ at the GS is denoted as $\tau_n$.%, where the weight of each edge $e\in\mathcal{E}$ is the distance between the end-vertices.

\subsection{Routing}
%\textcolor{blue}{Section for network or only some details here, of ISL RF}.
Routing is needed to forward the data from a satellite $n$, where the task is generated, towards a satellite $n_m$ that can download the data to a GS $m\in\mathcal{M}$. In this context, the objective of the routing algorithm is to select a path $\mathcal{S}_n$, from vertex $n\in \mathcal{N}$ in graph $\mathcal{G}$, defined as a set of ordered vertices
\begin{equation}
    \mathcal{S}_n = \left\{n, \ldots, n_m, m\right\}=\left\{s_n^{(1)}, \ldots, s_n^{(|\mathcal{S}_n|-1)}, s_n^{(|\mathcal{S}_n|)}\right\};
\end{equation}
hence, the $i$-th vertex in the path $\mathcal{S}_n$ is $s_n^{(i)}$.

Communication in both ISLs and DL takes place via RF unicast links, modeled as additive white Gaussian noise (AWGN) channels. Hence, the data rate for communication from the $i$-th to the $(i+1)$-th vertex in $\mathcal{S}_n$
   is calculated as
\begin{equation}
    R_i = B\log_2 \left( 1 + p_ih_i^2\right),
    \label{eq: shannon}
\end{equation}
where $B$ is the allocated bandwidth, $p_i$ is the transmission power used by the $i$-th satellite and $h_i^2$ is the respective squared channel coefficient normalized by the receive noise power.
%Note that the last hop between $m_n$ and GS is supported via direct downlink (DL) connection. 
%Connection through a relay consisting of NGEO-GEO-GS link is out of the scope of this study.
We consider that only the Earth surveillance data is transmitted by the satellites and, hence, there are no other traffic flows in the constellation. Furthermore, the LEOs are not shared between tasks due to stringent energy constraints. Besides, we assume all tasks to be programmed, so that they are generated simultaneously in the network. Building on this, the resulting routing latency for path $\mathcal{S}_n$ can be defined as
\begin{equation}
    T^{DL}_n =
    \sum_{i=1} ^{|\mathcal{S}_n|-1} \left(T^{comm}_i + T^{prop}_i\right)=
    \sum_{i=1} ^{|\mathcal{S}_n|-1} \left(\frac{L_n}{R_i} + T^{prop}_i\right),
    \label{eq: time_dl}
\end{equation}
where $T_i^{comm}$ is the communication time spent in the link between the $i$-th and $(i+1)$-th vertices in $\mathcal{S}_n$ and it is defined as the ratio between the data size $L_n$, and data rate $R_i$. $T^{prop}_i$ is the corresponding propagation delay.

%%%%%%%%%%%%%%%%%%%%%%%%%%%%%%%%%%%%%%%%%%%%%%%%%%%%%%
\subsection{Edge or Cloud Computing}
The processing of a task can be performed at the satellite $n$ (edge computing) where the task is generated or at the GS (cloud computing), whereas the intermediate satellites are only for forwarding.
 We assume a processing model that can encompasses either compression, fault detection or classification, hence, the tasks generated at different satellites may have different characteristics. Specifically, for a task generated at satellite $n$: $D_n$ is the original data size of the task and $F_n$ is the size after the processing (e.g., the result). If the satellite $n$ decides to process the task, the amount of data reduction due to processing is calculated as $\rho_n = {D_n}/{F_n} \geq 1$. 
%\begin{equation}
%    \rho_n = \frac{D_n}{F_n} \geq 1
%\label{eq:rho}
%\end{equation}
Note that these values may be different across the satellites depending on the characteristics of the tasks.
%Even though in practice there are several GS, they all will be envisioned under a unique node, which eases notation.
%The collaborative processing between satellites within an edge computing environment is left for future research. 

The decision between edge and cloud computing is defined by variable $l_n$, which takes the value of 1 when for edge computing and 0 for cloud computing. Building on this, the data size for a task generated at the $n$-th satellite is 
\begin{equation}
    L_n = l_nF_n + (1 - l_n)D_n = D_n\left(l_n\frac{1}{\rho_n} + (1 - l_n)\right)
    \label{eq: data_size}
\end{equation}

We consider a model to calculate the energy consumption of the satellites due to the processing of the task that captures the most relevant CPU parameters \cite{double_edge, multi_mec, 4c, survey, v_param}. In this model, the energy consumption per processing operation is proportional to the square of the clock frequency of the CPU $f_\text{CPU}$, in cycles per second, times a constant $\nu$, which is the effective capacitance coefficient of the processor~\cite{v_param}. Building on this, the energy consumption at the $n$-th LEO satellite due to processing is modeled as
\begin{equation}
    E_n^{proc} =l_nC_p(D_n) = l_nD_n z f_\text{CPU}^2\nu,
\label{eq: consumption_energy}
\end{equation}
where $C_p(D_n)$ is the energy consumption to process a task of size $D_n$, and $z$ is the number of CPU cycles to process 1 bit of data. %This model has been validated with experimental data from mobile devices \cite{v_experiment} and similar processors are considered for LEO satellites~\cite{esa_leo}.
%We do not need to distinguish between ISL and DL as they can be modelled individually with particular constraints in their data rates and distances.
%Observe that $T_i^{prop}$ is a function of the distance between the nodes, this is, the edge weight $w(e)$; conversely, $T_i^{comm}$ is a function of $w(e)$ along with information of the nodes (e.g., transmission power).
%In general, it is challenging describing the energy consumption of a processor. A more precise model depends on the number of instructions executed per task, the energy and the time spent per instruction. Nevertheless this parametrization is unmanageable because these specifications are complex to estimate: the number of instructions heavily depends on the algorithm implementation, and the energy and time per instruction depend on the instruction itself (e.g., type of memory access, CPU architecture, supply voltage, etc.).
Furthermore, we define the delay associated to processing the $D_n$ bits at vertex $n$ and at the GS, respectively, as
\begin{equation}
    T^{proc}_n =\frac{D_nz}{f_\text{CPU}};\qquad
    T^{proc}_m =\frac{D_nz}{kf_\text{CPU}}
    \label{eq: t_proc}
\end{equation}
We assume $k>1$ such that the processing at the GS is faster.
As stated in (\ref{eq: consumption_energy}) and (\ref{eq: t_proc}), the processing parameters are considered constant and identical for all processors. %Several research lines focus on optimizing these parameters in order to lead to convenient trade-offs (see \cite{survey} and references therein).

\subsection{Radio Resource Management}
The energy consumption due to the communication subsystem can be shaped by $C_{t}(L_n,R_i,p_i)$, that models the energy consumption of transmitting $L_n$ bits at data rate $R_i$ and with power $p_i$. A representative energy consumption model is
\begin{equation}
    C_{t}(L_n,R_i,p_i) = p_i\frac{L_n}{R_i}
\end{equation}

RF power amplifiers are a key component in satellite communications. % and its energy consumption cannot be neglected.
%For satellite RF communications, Traveling Wave Tube Amplifiers (TWTAs) are the main amplifier choice because they offer higher power at higher frequencies \cite{twta}.
From \cite{pa_model}, if we consider a narrow-band transmission, the power consumption at the amplifier can be linearly modelled as
\begin{equation}
    P_{c,i} = P_{fix} + \frac{c_0}{\eta}p_i,
\end{equation}
where $P_{fix}$ is the power consumption independent of the output power of the amplifier; $c_0$ is a scaling coefficient for the power loading dependency;
%, which depends on other subsystems such as the base band module, though these will not be considered
$\eta$ is the drain efficiency of the amplifier%, defined as the ratio between the output and consumed power by the amplifier
; $p_i$ is the output power of the amplifier, that is, the transmitted power. There is a maximum output power $P_{out}^{max}$ that limits the transmitted power of the amplifier. Consequently $p_i\in\left[0,P_{out}^{max}\right]$ for all $i$.
%Power amplifiers operate close to the saturation point, where the efficiency is maximum. The Output Back-Off (OBO) prevents the device from trespassing the linear region and determines the efficiency of the amplifier \cite{pa_model}. The OBO characterizes the drain efficiency of the amplifier.

The overall energy consumption model due to the RF subsystem including the power amplifier can be represented by defining $\mu=1+c_0/\eta$ as
\begin{equation}
     E_n^{RF} = \sum_{i=1} ^{|\mathcal{S}_n|-1} \left(P_{c,i} + p_i \right)\frac{L_n}{R_i}=
     \sum_{i=1} ^{|\mathcal{S}_n|-1} \left(P_{fix} + \mu p_i\right)\frac{L_n}{R_i},
\end{equation}

\section{Energy-Efficient Joint  Satellite  Computation and  Communication (Sat2C) Optimization}
The objective to be minimized is the total energy per bit, this is, the sum energy over all paths in transmission. Considering the definition of $\rho_n$,
%the data size $D_n$ is a common factor to all terms, in such a way that
the energy per bit can be expressed as
\begin{align}
    E_b^{total}=&\sum_{n\in\mathcal{N}} E_{b,n}^{total}= \sum_{n\in\mathcal{N}} \frac{1}{D_n} E_n^{total}
    \\=&\sum_{n\in\mathcal{N}}\frac{1}{D_n} \left(E_n^{RF} + E_n^{proc}\right)
    = \sum_{n\in\mathcal{N}} E_{b,n}^{RF} + E_{b,n}^{proc}\nonumber
\end{align}
$E_{b,n}^{RF}$ and $E_{b,n}^{proc}$ are the energy per communicated and processed bit at the $n$-th satellite, respectively. With that, we propose the following optimization problem that merges routing, transmit power allocation and processing task decision:
\begin{equation*}
\begin{aligned}
& \underset{\left\{\mathcal{S}_n, \{p_{i}\}_{s_n^{(i)}\in\mathcal{S}_n}, l_n\right\}_{n\in\mathcal{N}}}{\text{minimize}} & & E_b^{total} && \textbf{(P1)} \\
& \hspace{7pt} \qquad\text{subject to}              & &  \mathcal{S}_n\cap\mathcal{S}_{n'}\in\mathcal{M}\cup\emptyset,\, \forall n\neq n' && \text{(C1)} \\
&                          & &  s_n^{(1)}\in\mathcal{N} && \text{(C2)}\\
&                          & &  s_n^{(|\mathcal{S}_n|)}\in\mathcal{M} && \text{(C3)}\\
&                          & &  g_i\frac{E^{proc}_n}{T^{proc}_n} + p_i\leq P_i,\; \forall s_n^{(i)}&& \text{(C4)}\\
&                          & &  p_i\leq P_{out}^{max},\; \forall s_n^{(i)} && \text{(C5)}\\
&                          & &  T_n\leq \tau_n && \text{(C6)}\\
&                          & &  l_n\in\{0,1\} && \text{(C7)}
\end{aligned}
\label{opeq}
\end{equation*}
for all $n$ in all constraints. Constraint (C1) ensures path deconfliction, that is, that any two paths share no intermediate nodes, whereas (C2) and (C3) ensure feasibility, that is, that each path starts at the origin satellite and terminates at a GS, respectively. Constraint (C4) limits the power consumption in processing and communication to be below the available at the satellite payload, $P_i$. Parameter $g_i$ is predefined and takes the value 1 when $i=1$, and 0 otherwise. Notice this constraint evinces the joint processing and communication subsystems power budget. Likewise (C5) restricts the transmitted power to be below the maximum available at the amplifier. Constraint (C6) refers to the total latency:
\begin{align}
T_n &= T^{DL}_n + l_nT^{proc}_n +(1-l_n)T^{proc}_m 
\end{align}
%In light of \cite{federated} we will analyse the problem in terms of the communication time.
Problem (P1) is NP and cannot be solved optimally. Therefore, we first tackle the routing problem and, for the preset paths, we optimally solve the power allocation and offloading decisions problem.

\subsection{Routing Procedure}
For notation simplicity, we drop the superscript in $T^{comm}_i$ to simply $T_i$. Considering the definition of $T_i$ used in (\ref{eq: time_dl}) and (\ref{eq: shannon}), we can express the communication power as
\begin{align}
p_i=\frac{1}{h_i^2}\left( 2^{\frac{L_n}{B T_i}} -1\right)
\end{align}
In this way, the energy cost function becomes
\begin{multline}
E_b^{total} = \sum_{n=1}^N \frac{1}{D_n} \Bigg( \sum_{i=1} ^{|\mathcal{S}_n|-1}\left(P_{fix} + \frac{\mu}{h_i^2}\left( 2^{\frac{L_n}{B T_i}} -1\right)\right)T_i\\+ l_nC_p(D_n) \Bigg)
\label{eq:cost}
\end{multline}
Besides increasing the power consumption towards a more realistic model, the power amplifier changes the convexity of the problem: for $P_{fix}\neq0$, the objective (\ref{eq:cost}) is convex and non-monotonic in $T_i$; whereas, without considering the power amplifier model, the energy function is decreasing in $T_i$. These two results are known outcomes in the literature of energy efficiency for wireless networks \cite{fractional_program, olga}. %From this, the minimum energy is obtained with inequality in the (C6) constraint.

% \begin{lemma}
% For $P_{fix}\neq0$, the objective (\ref{eq:cost}) is convex and non-monotonic in $T_i$.
% \label{lemma:monotone}
% \end{lemma}

%Minimizing the propagation time is a common practice in routing procedures and it is compliant with the energy-efficient framework.

%Nevertheless, the problem is more complex, as a different route implies different channels and, thus, a modification in the energy function. However, we choose minimizing the propagation time as the routing metric since it is widely extended heuristic procedure.

\begin{figure}[bt]
\includegraphics[width=\columnwidth]{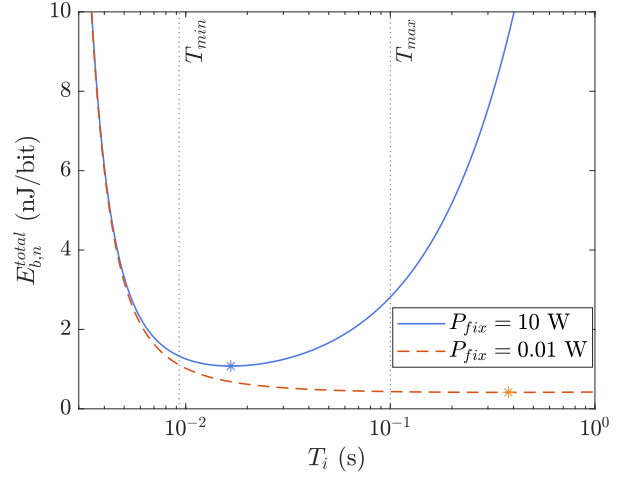}
\centering
\caption{Problem (P1) particularized for a unique ISL and for two different values of $P_{fix}$. The stars mark the minimums.}
\label{fig:energy_regions}
\end{figure}

%Routing via minimization of the propagation delay is equivalent to using path-loss: the lower the delay, the smaller the distance and the smaller the path-loss.
%Besides, according to \cite{pathloss}, multirate routing in LEO constellations via path-loss provides a smaller overall latency and favours paths with large data rates in contrast to other metrics.

%The propagation time plays an important role in the optimization because it does not only affect constraint (C6), but it also alters the channel coefficients in the energy function depending on the selected path.
%However, for two different power amplifier configurations, its minimization leads to minimum energy consumption. 
Figure \ref{fig:energy_regions} exhibits (P1) particularized for a unique ISL, this is, for $n=1$ and $|\mathcal{S}_n|=2$. It is expressed in $T_i$, for two different values of $P_{fix}$ and we assume $l_n=0$. The upper bound, $T_{max}$, corresponds to transmitting at minimum power (i.e., $\tau_n$ minus the propagation time), whereas the lower bound, $T_{min}$, corresponds to transmitting at maximum power. For $P_{fix}=0.01$ W, the minimum (star) cannot be achieved unless the upper bound is pushed to the right. This can be attained by reducing the propagation time, this is, using shorter routes so that $T_{max}$ increases.

Nevertheless, even though routing via minimum propagation time is a common practice, it is not straightforward to generalize the previous result for multiple hops. Thus, we will follow this heuristic procedure and leave the optimality of this routing  strategy in P1 for future work.
%Therefore, even though the minimum energy may not be achieved with the shortest route, the heuristic procedure of routing via the propagation time ensures that more energy can be devoted to communication.

We define the optimization problem (P2) to determine the sets $\mathcal{S}_n$ as the paths minimizing the total propagation time.
\begin{equation*}
\begin{aligned}
& \underset{\left\{\mathcal{S}_n\right\}_{n=1}^N}{\text{minimize}} & & \sum_{n=1}^N \sum_{i=1} ^{|\mathcal{S}_n|-1}T^{prop}_i && \textbf{(P2)} \\
& \text{subject to}              & &  \mathcal{S}_n\cap\mathcal{S}_{n'}\in\mathcal{M}\cup\emptyset,\, \forall n\neq n'  && \text{(C1)} \\
%&                          & &  \mathcal{S}_n\cap\mathcal{A}=\{\mathcal{A}(n), GS\} & n=1,\dots,N & & \text{(C2)}\\
&  & &  s_n^{(1)}\in\mathcal{N}(n),\; \forall n && \text{(C2)}\\
&  & & s_n^{(|\mathcal{S}_n|)}\in\mathcal{M},\; \forall n && \text{(C3)}
\end{aligned}
\label{opeq}
\end{equation*}
%We assume that $\tau_n$ is much larger (e.g., some order of magnitude) than any propagation time so that all solutions remain feasible.
Finding the optimal solution to the multiple shortest path problem with constraints is NP-hard. There is up-to-date literature dealing with path search for multiple agents with node deconfliction \cite{path_deconfliction}. %This has been tackled through relaxation methods to apply convex optimization \cite{relaxation} and via tailored algorithms \cite{canine}.
%The development of an specific algorithm for solving (P2) is out of the scope of this study. In contrast, we have proposed a set of algorithms that fit within the energy minimization framework, this is, based on the minimization of the propagation time. Moreover, our framework decouples the routing from the other tasks, which allows to substitute it for any alternative algorithm.
The development of a competitive algorithm is out of the scope of this paper. Our aim is to show that (P2) leads to an energy-efficient procedure, even when decoupling the routing problem is suboptimal.
%It allows to study specific algorithms for routing.

%We introduce the original SHIELD algorithm (\textbf{S}ubmodular \textbf{HIE}rarchica\textbf{L} \textbf{D}ijkstra's algorithm) in Algorithm \ref{alg: shield}, used for routing. In practice it is not necessary to compute $\mathcal{S}$, this is, all feasible paths from $\mathcal{N}$ to the GS. For a given node, only the shortest path until the GS minimizes the marginal gain, meaning that we only need to compute the shortest path for each feasible node at every iteration. We use the Dijkstra's algorithm, which is a well-known algorithm used to efficiently compute the shortest path between two nodes in a graph \cite{dijkstra}.

%Since the SHIELD algorithm runs the Dijkstra's algorithm $N^2$ times, its time complexity can be bounded by $\mathcal{O}(N^2(\mathcal{|V|}+\mathcal{|E|}log\mathcal{|V|}))$, where $\mathcal{|V|}$ and $\mathcal{|E|}$ are the number of vertices and edges in the graph, respectively.

% \begin{algorithm}[t]
% \SetAlgoLined
% \KwIn{$\mathcal{G}$ and $\mathcal{N}$}
% \KwOut{$\mathcal{S}_i \text{ for } i=1,\dots,N$}
% %\KwResult{$\mathcal{S}_n, n=1,\dots,N$}
% \vspace{4 pt}
% Initialize $\mathcal{S}_i=\varnothing\ \forall i$\;

%  \For{$i=1,\dots,N$}{
%     \For{$j=i,\dots,N$}{
%         $e_j = \text{Dijkstra}(\mathcal{G},\mathcal{N}(j), GS)$\;
        
%         $j^*=\text{argmin}_j\  f_j(\mathcal{S}_j\cup \{e\})$
%     }
%     $\mathcal{S}_i\gets \{e_{j^*}\}$\;
    
%     $\mathcal{G}\gets \{\mathcal{G}\setminus\mathcal{S}_i\}\cup\{GS\}$
%  }
%  \caption{SHIELD}
%  \label{alg: shield}
% \end{algorithm}

\subsection{Radio Resource Allocation and Offloading Strategy}
As a result of the routing procedure, the RM problem is decoupled for every path $\mathcal{S}_n$. We focus on the transmit power problem, because the low dimensionality of $l_n$ allows to solve the former for $l_n=\{0,1\}$ and choose the decision minimizing the energy for every path.
When particularizing (P1) for every $\mathcal{S}_n$, we observe that it can be cast as a maximization fractional program:
\begin{equation*}
\begin{aligned}
& \underset{\{p_{i}\}_{s_n^{(i)}\in\mathcal{S}_n}}{\text{maximize}} & & -E_{b,n}^{total} && \textbf{(P3)} \\
& \hspace{-19pt}\qquad\text{subject to}              & &  g_i\frac{E^{proc}_n}{T^{proc}_n} + p_i\leq P_i,\; \forall s_n^{(i)} && \text{(C1)} \\
&                          & &  p_i\leq P_{out}^{max},\; \forall s_n^{(i)} && \text{(C2)}\\
&                          & &  T_n\leq \tau_n && \text{(C3)}
\end{aligned}
\label{op_frac}
\end{equation*}
%Parameter $\gamma_n$ corresponds to the communication time only, this is, $\tau_n$ minus the processing and propagation delays.
Problem (P3) is a Sum of Ratios Problem (SoRP) \cite{fractional_program}, as the cost function can be rewritten as (\ref{eq:cost_SoRP}) %, in which every numerator and denominator are concave and convex, respectively,
and the constraints are convex in $p_i$. This SoRP can be solved optimally with the Dinkelbach’s algorithm because all terms are decoupled for every $p_i$. We call the overall optimization framework Sat2C, and it is described in Algorithm \ref{alg: Sat2C}.
%\textcolor{blue}{(Should we clarify that this is not optimal because we break down the overall problem into routing and RRM? No need to discuss about convergence, right?)}
\begin{align}
-E_{b,n}^{total}=\sum_{i=1} ^{|\mathcal{S}_n|-1}\frac{P_{fix}+p_i}{\log_2\left(\frac{1}{1+p_ih_i^2}\right)}
\label{eq:cost_SoRP}
\end{align}

\begin{algorithm}[t]
\SetAlgoLined
\KwIn{$\mathcal{G}$ and $\mathcal{N}$}
\KwOut{$\mathcal{S}_n,\, l_n,\, \{p_{i}\}_{{s_n^{(i)}\in\mathcal{S}_n}} \; \text{ for } n\in\mathcal{N}$}
%\KwResult{$\mathcal{S}_n, n=1,\dots,N$}
\vspace*{4 pt}
$\{\mathcal{S}_n\}_{n\in \mathcal{N}}\gets \text{solve (P2)}$\;
\vspace*{2pt}
 \For{$n\in \mathcal{N}$}{%\For{$n=1,\dots,N$}{
    $E_{b,n}^{total\{0\}}, \{p_i^{\{0\}}\}_{s_n^{(i)}\in \mathcal{S}_n}\gets \text{ solve (P3) for $l_n=0$}$\;
    \vspace*{2pt}
    $E_{b,n}^{total{\{1\}}},\{p_i^{\{1\}}\}_{{s_n^{(i)}\in\mathcal{S}_n}}\gets \text{ solve (P3) for $l_n=1$}$\;
    \vspace*{2pt}
    $l_n\gets \text{argmin}_{l_n}\{E_{b,n}^{total{\{l_n\}}}\}$\;
    \vspace*{2pt}
    $\{p_i\}_{{s_n^{(i)}\in\mathcal{S}_n}}\gets \{p_i^{\{l_n\}}\}_{{s_n^{(i)}\in\mathcal{S}_n}}$\;
 }
 \caption{Sat2C}
 \label{alg: Sat2C}
\end{algorithm}

\section{Results}
Next we evaluate Sat2C with shortest path routing. As to the routing, we have implemented a greedy algorithm based on submodularity that minimizes the propagation delay. This provides solutions that are scalable, Pareto efficient and with one-half-approximation guarantee \cite{submodularity}.

\subsection{Setup}

\begin{table}[t]
    \centering
    \caption{Communication parameters for the Kepler constellation}
    \begin{tabular}{@{}llll@{}}
         \toprule
         Parameter & LEO to GS
         &ISL  \\\midrule
         Carrier frequency (GHz) & $20$ & $26$\\
         Bandwidth (MHz) & $500$ & $500$\\
         Maximum transmission power (W) & $10$ & $10$\\
         Antenna diameter (Tx -- Rx) (m) & ($0.26$ -- $0.33$) & ($0.26$ -- $0.26$)\\
         Antenna gain (Tx -- Rx) (dB) & ($32.13$ -- $34.20$) & ($34.41$ -- $34.41$)\\
         Pointing loss  (Tx -- Rx) (dB) & ($0.3$ -- $0.3$) & ($0.3$ -- $0.3$)\\
         Antenna efficiency (Tx -- Rx) (--) & ($0.55$ --  $0.55$) & ($0.55$ -- $0.55$) \\
         Noise temperature (K) & $50$ & $290$\\
         Noise figure (dB) & $1.5$ & $2$\\
         Noise power (dB) & $-119.32$ & $-114.99$\\
         \bottomrule
    \end{tabular}
    \label{tab:Kepler_comm_params2}
\end{table}

We consider a Walker star constellation, with $7$ orbital planes with $20$ LEO satellites per plane. The orbital planes correspond to polar orbits and are deployed at the same altitude of $H=600$ km. There are 26 GSs, according to the KSAT infrastructure \cite{ksat}. There are $N=10$ randomly selected source satellites uniformly distributed across the globe.

%The communication parameters are adjusted according to the spectral efficiency for the DVB-S2X system \cite{dvb}: the modulation and coding schemes cover the range of SNRs $[-2.85, 19.57]$\,dB, such that the spectral efficiency is $0$ if $\text{SNR}<-2.85$\,dB and $5.90$ if $\text{SNR}\geq19.57$\,dB. Hence, the antenna design and the selection of transmission power is focused on achieving SNRs within this interval. The communication system is located in the Ka-band and Table \ref{tab:Kepler_comm_params2} lists a configuration of parameters satisfying the above requirements.
The antenna design and the maximum transmission power are adjusted according to the spectral efficiency for the DVB-S2X system \cite{dvb}. Table \ref{tab:Kepler_comm_params2} lists a configuration of parameters satisfying the above requirements.%We assume clear sky conditions.

Regarding the power amplifier subsystem, we set $\eta=0.65$ \cite{obo}, $c_0=\frac{\pi}{4}\eta$ \cite{pa_model}, $P_{fix}=5$ W and $P_{out}^{max}=10$ W. With respect to the processing, $f_\text{CPU}=250$ MHz, $k=4$, $z=737.5$ CPU cycles/bit and $\nu=10^{-27}$ J/Hz$^3$ \cite{4c, double_edge}.
We set $D_n=1.2$ Mb,
%which results in a processing time below 4 seconds
$\tau_n=10$ s, and $\rho_n=4\text{ for all } n$.

We use a Monte Carlo setup of 1000 experiments and average the results over all runs and routes.

\begin{figure}[bt]
\vspace*{-0.1cm}
\hspace*{-0.38cm}
\includegraphics[width=1.05\columnwidth]{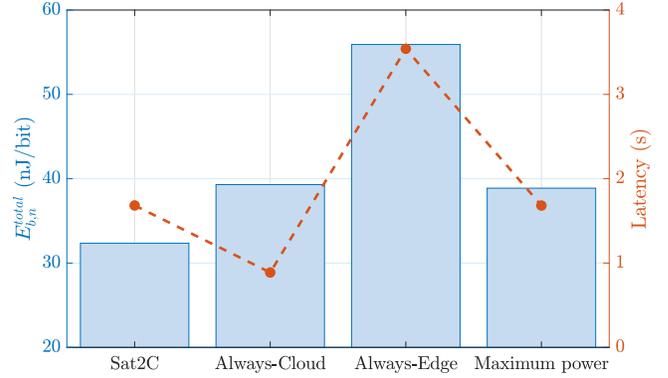}
\centering
\vspace*{-0.5cm}
\caption{Benchmark comparison of the mean-path energy per bit (bar plot) and latency (markers).}
\label{fig:energy_results}
\end{figure}

\subsection{Sat2C Performance}
The performance of Sat2C is compared to three alternative approaches:
\begin{itemize}
    \item \textbf{Always-Cloud policy}: optimal transmission powers and always offloading the task. It is equivalent to Sat2C with $l_n=0 \text{ for all } n$.
    \item \textbf{Always-Edge policy}: optimal transmission powers and never offloading the task. It is equivalent to Sat2C with $l_n=1 \text{ for all } n$.
    \item \textbf{Maximum Power policy}: maximum transmission power and computes the optimal task allocation decision for each route.
\end{itemize}

Figure \ref{fig:energy_results} displays the mean total energy per bit (bar plot) and the corresponding mean total latency (markers). These results evince the importance of routing in LEO constellations, as the latency is reduced in several orders of magnitude. Sat2C outperforms the other alternatives, offering a proper trade-off between energy and latency, which demonstrates the importance of jointly optimizing resources and computation decisions.%For instance, the Always-Cloud policy achieves the shortest latency because the processor at the GS is faster, even though it increases the energy consumption due to communication.
%Similarly, always offloading the tasks to the GS provides minimum delay at the expenses of consuming more energy. Sat2C provides a suitable trade-off between energy and latency. Notice it takes the same delay as the Maximum Power policy, but consumes less energy due to the optimal RRM procedure.

\subsection{Parametric Analysis}
Since these results highly depend on the CPU specifications and the nature of the task, in the following we analyse the effect of these parameters, that help to dimension the system.

\begin{figure}[bt]
\includegraphics[width=\columnwidth]{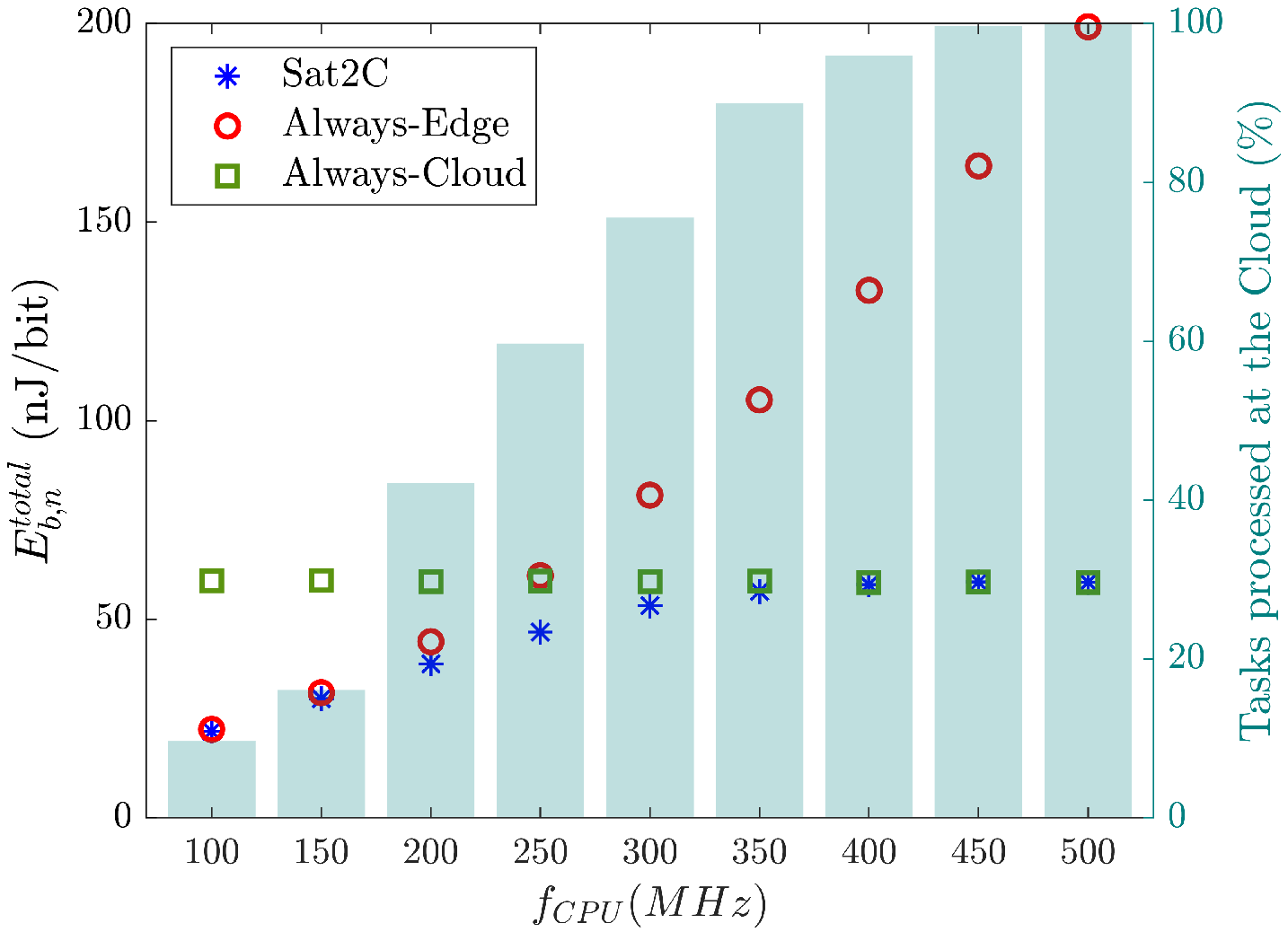}
\centering
\vspace*{-15pt}
\caption{Effect of $f_\text{CPU}$ in $E_{b,n}^{total}$ (markers) and the percentage of tasks offloaded by Sat2C (bar plot).}
\label{fig:f_cpu}
\end{figure}

As $f_\text{CPU}$ increases, the energy spent in processing does as well, suggesting that a faster processor does not suit energy minimization. In Figure \ref{fig:f_cpu}, the Always-Edge policy is heavily affected by the speed of the CPU, whereas the Always-Cloud algorithm is not affected by the computation model. The optimal solution of Sat2C approaches the Always-Edge strategy when it is cheap to compute locally (i.e., low frequency) and to the Always-Cloud policy at high frequency. In between, only some tasks are computed on the edge devices.

\begin{figure}[bt]
\vspace*{-10pt}
\includegraphics[width=\columnwidth]{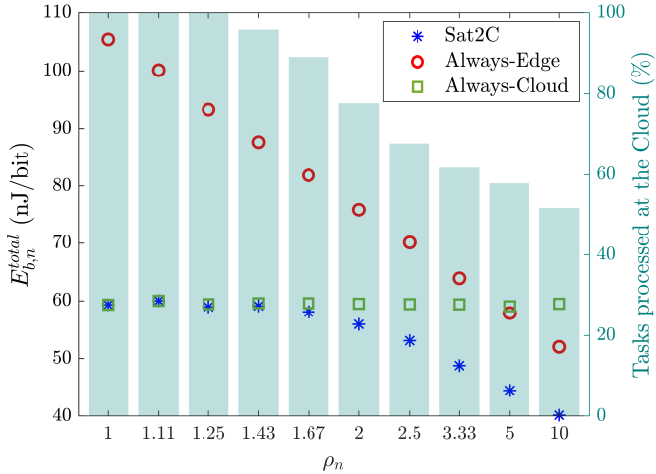}
\centering
\vspace*{-15pt}
\caption{Effect of $\rho_n$ in $E_{b,n}^{total}$ (markers) and the percentage of tasks offloaded by Sat2C (bar plot).}
\label{fig:rho_cpu}
\end{figure}

In Figure \ref{fig:rho_cpu}, the Always-Edge policy decreases with $\rho_n$ because the more compressed is the data, less bits are transmitted. Thus, the Always-Edge policy shows how the communication energy is reduced as the processing compresses the information. Sat2C degrades to the Always-Cloud policy when the data is not compressed, as the satellite would spend energy processing to transmit the same amount of data.

\section{Conclusions}
In this paper we propose a novel problem of energy-efficient satellite edge computing that jointly optimizes transmit powers, routes and computing decisions.
We develop an algorithm that decouples the routing from the RM problem. While the former remains NP, the latter can be solved optimally for the established paths. Experimental results show that this algorithm outperforms the baseline store-and-forward policy and provides a suitable trade-off between energy consumption and latency. The simulations highlight the importance of routing to meet the service time requirements and the relevance of the CPU parameters in the dimension of the system. In future work we will consider a more generic offloading strategy based on partial offloading, so that the network can support larger amounts of data and more demanding tasks.

\section{Acknowledgement}
We thank Nader Alagha, Alberto Mengali, and Petar Popovski for providing insight and expertise that assisted the research, and comments that greatly improved the manuscript.

This work has been performed under a programme of and funded by the European Space Agency. The view expressed herein can in no way be taken to reflect the official opinion of the European Space Agency.

\bibliographystyle{ieeetr}
\bibliography{refs}

\end{document}